\documentclass[aps,prl,twocolumn,showpacs,amsmath,amssymb]{revtex4}
\usepackage{dcolumn}
\usepackage{bm}
\usepackage{graphicx}
\usepackage{subfigure}
\usepackage{epstopdf}
\usepackage[makeroom]{cancel}

\renewcommand{\section}[1]{{\par\it #1.---}\ignorespaces}

\begin{document}
\title{Majorana fermions at odd junctions in a wire network of ferromagnetic impurities}
\author{Kristofer Bj\"{o}rnson and Annica M. Black-Schaffer}
\affiliation{Department of Physics and Astronomy, Uppsala University, Box 516, S-751 20 Uppsala, Sweden}

\begin{abstract}
We consider a wire network of ferromagnetic impurities on the surface of an $s$-wave superconductor with strong Rashba spin-orbit interaction.
Within the topological phase, zero-energy Majorana fermions appear at wire end-points as well as at junctions between an odd number of wire segments, while no low-energy states are present at junctions between an even number of wire segments, providing strong experimentally accessible signatures for Majorana fermions. We also investigate the quasiparticle energy gap with respect to varying the Rashba spin-orbit coupling and magnetic impurity strength.

\end{abstract}

\maketitle

During the past few years, topological superconductivity and the often accompanying Majorana fermions have attracted a significant amount of interest~\cite{PhysUsp.44.131, PhysRevLett.95.226801, PhysRevLett.96.106802, Science.314.1757, Science.318.766, Nature.452.970, NatPhys.5.398, NatPhys.5.438, Science.325.178, RevModPhys.82.3045, RevModPhys.83.1057, PhysRevLett.100.096407}.
The interest is at least two-fold.
First of all, topological materials, and Majorana fermions in particular, have a fundamental appeal because of analogies to phenomena in high-energy physics.
A key feature of topological materials is the topologically protected Dirac spectrum that appears at their boundaries, which resembles the spectrum of relativistic particles~\cite{TheUniverseInAHeliumDroplet}.
Topological superconductors also offers the promise of finding particles so far only theoretically predicted in high-energy physics, the so-called Majorana fermions, which are the solutions to a real-valued Dirac equation, and thus are their own antiparticles~\cite{NatPhys.5.614}.
Secondly, condensed matter realizations of Majorana fermions have also been proposed to be used to construct robust qubits for quantum computing, generating significant interest also from an applied point of view~\cite{PhysUsp.44.131, Science.339.1179}.

One prominent system where theory predicts that condensed matter realizations of Majorana fermions appear is at the end points of a wire of ferromagnetic atoms deposited on top of an $s$-wave superconductor with strong Rashba spin-orbit interaction~\cite{PhysRevLett.103.020401, PhysRevB.82.134521, PhysRevB.88.020407}, or the physically equivalent setup with magnetic atoms forming a helical magnetic state \cite{PhysRevB.84.195442,PhysRevLett.111.147202, PhysRevLett.111.186805, PhysRevLett.111.206802, PhysRevB.88.155420}.		
Recent scanning tunneling microscopy and spectroscopy (STM/STS) experiments on such a system composed of magnetic Fe atoms deposited on Pb, which is a conventional $s$-wave superconductor with Rashba spin-orbit interaction, have observed the first hallmark of Majorana fermions; zero-energy bias peaks in the local density of states (LDOS) at the wire end points~\cite{Science.346.602, arXiv.1505.06078, PhysRevLett.115.197204}.
However, zero-energy states can have multiple origins and might very well not be Majorana fermions~\cite{PhysRevB.91.094505, PhysRevLett.115.127003}. Thus, stronger evidence is needed in order to rule out other explanations.

In this work we propose to build a network of wires by using STM to deposit ferromagnetic impurity atoms on a Rashba spin-orbit coupled superconductor and find signatures that are clearly distinct to Majorana fermion zero-energy states.
We show through realistic numerical calculations that well-localized Majorana fermions appear not only at end points of individual wires, but also at junctions between an odd number of wires. However, at junctions with an even number of wires, there are {\it no} low-lying energy states.

Networks of one-dimensional (1D) topological wires have previously been studied for future Majorana fermion braiding~\cite{NatPhys.7.412, PhysRevB.84.035120, PhysRevB.85.144501, NewJPhys.14.035019, PhysRevB.87.035113, PhysRevB.88.035121, NatNano.8.859, arXiv.1511.05153}, but we show that even the plain experimental observation of localized zero-energy states at odd junctions, together with the absence of such states at even junctions, directly provides strong evidence for that the zero-energy states corresponds to Majorana fermions.
We further investigate how the energy of the quasiparticle impurity bands depends on all parameters entering the system: Rashba spin-orbit interaction, magnetic impurity term, chemical potential, and superconducting order parameter. 
We verify that in a broad parameter range, there is a large difference in energy between Majorana fermions at odd junctions and the lowest-energy quasiparticle impurity states in even junctions.
This reaffirms a wire network as a very promising tool to probe the Majorana character of zero-energy states.
Moreover, the quasiparticle impurity energies also sets the excitation gap protecting the Majorana fermions from quasiparticle poisoning in odd junctions.
In particular, we show that a large Rashba spin-orbit interaction is highly beneficial for a large excitation gap.

\section{Model}
\begin{figure}
\includegraphics[width=140pt]{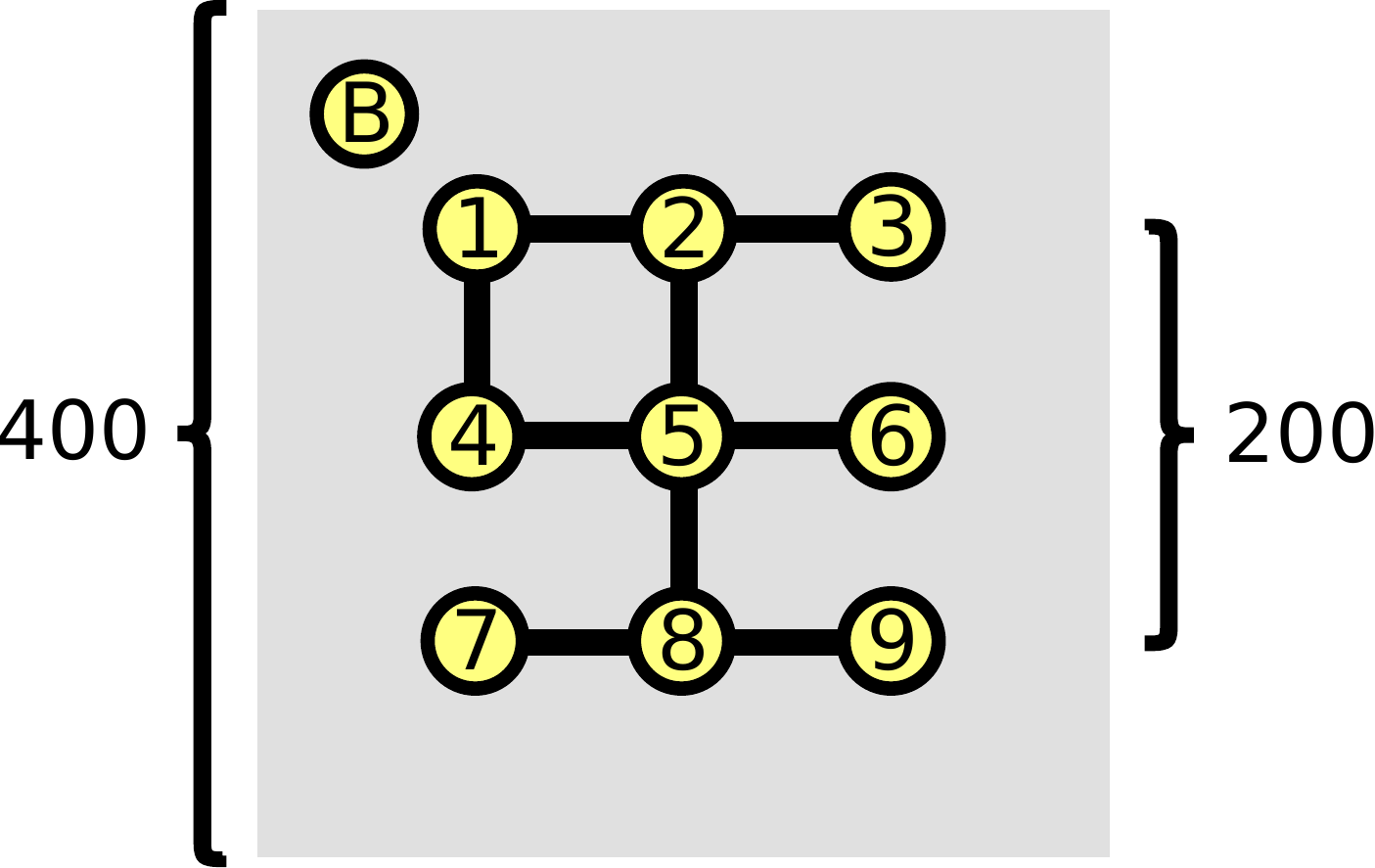}
\caption{Wire network of ferromagnetic impurities embedded in a conventional 2D superconducting layer with Rashba spin-orbit interaction. The total lattice size is $400\times 400$ sites, while each one-atom wide wire segment is $100$ sites long. Sites 1-9 represents all wire end and junction points, while B is a generic bulk site. }
\label{Figure:WireNetwork}
\end{figure}

The presence of ferromagnetic impurity atoms deposited on top of an $s$-wave superconductor with strong Rashba spin-orbit interaction~\cite{Science.346.602, arXiv.1505.06078, PhysRevLett.115.197204}, gives rise to the following effective Hamiltonian~\cite{PhysRevLett.103.020401, PhysRevB.82.134521, PhysRevB.88.024501, PhysRevB.84.180509, PhysRevB.91.214514, PhysRevLett.115.116602, PhysRevB.92.214501}
\begin{align}
	\mathcal{H} &= \mathcal{H}_{kin} + \mathcal{H}_{so} + \mathcal{H}_{sc} + \mathcal{H}_{V_z},
\label{Equation:Tight_binding_Hamiltonian} \\ 
	\mathcal{H}_{kin} &= -t\sum_{\langle\mathbf{i},\mathbf{j}\rangle,\sigma}c_{\mathbf{i}\sigma}^{\dagger}c_{\mathbf{j}\sigma} - \mu\sum_{\mathbf{i},\sigma}c_{\mathbf{i}\sigma}^{\dagger}c_{\mathbf{i}\sigma}, \nonumber \\ 
	\mathcal{H}_{so} &= \alpha\sum_{\mathbf{i}\mathbf{b}}\left(e^{i\theta_{\mathbf{b}}}c_{\mathbf{i}+\mathbf{b}\downarrow}^{\dagger}c_{\mathbf{i}\uparrow} + {\rm H.c.}\right), \nonumber \\ 
	\mathcal{H}_{sc} &= \sum_{\mathbf{i}}\left(\Delta c_{\mathbf{i}\uparrow}^{\dagger}c_{\mathbf{i}\downarrow}^{\dagger} + {\rm H.c.}\right), \nonumber \\
	\mathcal{H}_{V_z} &= -\sum_{\mathbf{i},\sigma,\sigma'}V_z(\mathbf{i})\left(\sigma_z\right)_{\sigma\sigma'}c_{\mathbf{i}\sigma}^{\dagger}c_{\mathbf{i}\sigma'}, \nonumber
\end{align}
modelled on a 2D square lattice.
Here $c_{\mathbf{i}\sigma}^{\dagger}$ ($c_{\mathbf{i}\sigma}$) is a creation (annihilation) operator for a $\sigma$-spin on site $\mathbf{i}$. Further, $\langle\mathbf{i},\mathbf{j}\rangle$ denotes summation over nearest neighbor indices, while $\mathbf{b}$ runs over all vectors pointing along the nearest neighbor bonds and $\theta_{\mathbf{b}}$ denotes the polar coordinate of this vector.
The superconductor is here represented by a generic band structure set by the hopping parameter $t$, chemical potential $\mu$, and spin-orbit coupling $\alpha$, while the superconducting $s$-wave order parameter is $\Delta$. For simplicity we measure energy in units of $t = 1$.
The presence of the magnetic impurity atoms is modelled by a Zeeman spin-splitting $V_{z}(\mathbf{i})$ being induced in the superconductor, which is only finite at the impurity sites, indicated in Fig.~\ref{Figure:WireNetwork}. An impurity wire enters a topologically non-trivial phase, hosting end-point Majorana fermions, for $V_z$ larger than some critical value~\cite{PhysRevB.92.214501}. Adatoms forming a helical magnetic state can be reduced to the same model, with the helicity of the adatoms generating the  spin-orbit coupling \cite{PhysRevLett.111.186805, PhysRevLett.111.206802}.
We are here primarily concerned with the directly experimentally measurable LDOS which we calculate using
\begin{align}
	\rho_{\mathbf{i}}(E) =& -\frac{1}{\pi}\sum_{\sigma}\textrm{Im}\left(G_{\mathbf{i}\mathbf{i}}^{\sigma\sigma}(E)\right),
\end{align}
following the Chebyshev-Bogoliubov-de Gennes method outlined in Ref.~[\onlinecite{PhysRevLett.105.167006}], where we truncate the Chebyshev expansion at $10^4$ coefficients. This method allow us to study wires with a realistic length that are embedded in a large superconducting system. Spin-polarized LDOS is calculated by ignoring the summation over $\sigma$ in the above expression.

\section{Majorana fermions}
\begin{figure}[tbh]
\includegraphics[width=225pt]{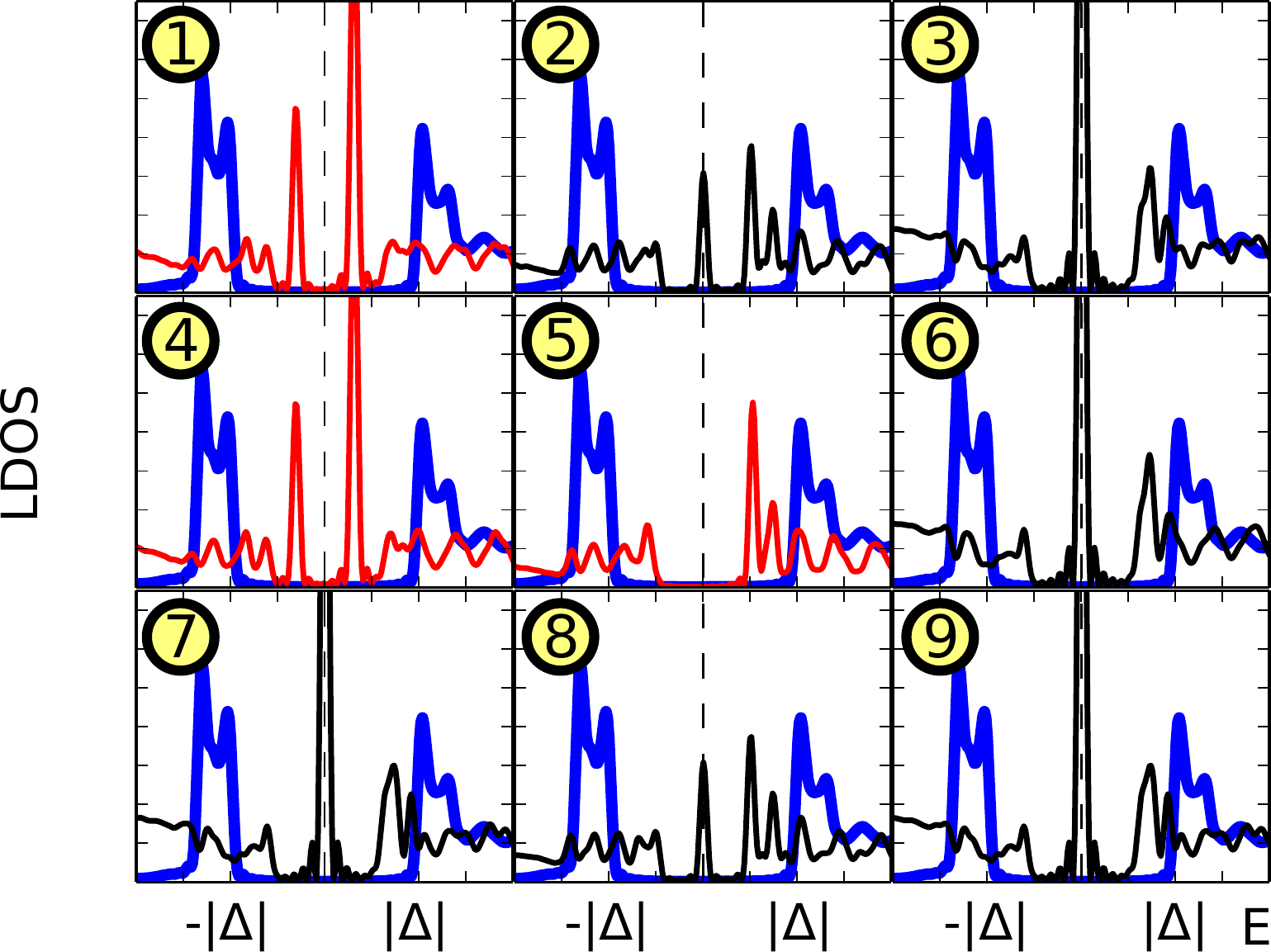}
\caption{LDOS on the sites marked 1-9 in Fig.~\ref{Figure:WireNetwork} (thin lines) and on the bulk site B (thick blue lines).
Zero-energy peaks are seen on sites 2, 3, and 6-9 (black), while such peaks are absent on sites 1, 4, and 5 (red).
The parameters are here $\mu = -4$, $\alpha = 0.3$, $\Delta = 0.1$, $V_z = 1.5$, which puts the wires comfortably within the topological phase.}
\label{Figure:LDOS}
\end{figure}

In Fig.~\ref{Figure:LDOS} we show the LDOS calculated on the sites marked by B (blue thick line) and 1-9 in Fig.~\ref{Figure:WireNetwork}.
Clear zero-energy peaks are observed on site 2, 3, and 6-9 (black lines).
All of these corresponds to junctions with an odd number of wire segments or single wire end points, henceforth named odd junctions.
In contrast, no zero energy peaks are observed on sites 1, 4, and 5 (red lines), which corresponds to junctions with an even number of wire segments, or simply even junctions.
This clear-cut distinction between even and odd junctions can be understood from the requirement that an even number of Majorana fermions have to be present in any condensed matter system:
First, consider a system with a single multi-wire junction.
In this case the number of wire end-points is even or odd, depending on whether the junction is even or odd. Thus for an odd junction one Majorana fermion has to appear somewhere else than at the end-points.
That this remaining Majorana fermion appears at the junction is clear when considering the junction as being created by bringing several end-points together. Then pairs of Majorana fermions are able to hybridize and can thus split in energy at the junction. In total, this leaves at least one unpaired Majorana at the junction whenever the number of incoming wires is odd.
Now, as long as the wire segments are long enough, the local physics at a certain junction should not depend on whether an incoming wire segment truly terminates in a single end-point or in another junction.
It is therefore expected that Majorana fermions appear at odd junctions in a wire network. 
In addition, Fig.~\ref{Figure:LDOS} shows that at even junctions, both 2- and 4-wire junctions, the original wire end-point states significantly hybridize, leaving no low-lying energy states and not even any states notably separated from the remaining quasiparticle impurity bands.

Next we note that the zero-energy peaks at site 2 and 8, corresponding to junctions between three wire segments, are notably smaller than the peaks at wire end-points.
Clearly, Majorana fermions at wire end-points can only spread along the wire network in a single direction, while the Majorana fermions at three-wire junctions can spread along all three directions.
Smaller peaks are therefore natural due to the Majorana fermions being localized over a larger region.
Also, as can be seen from Fig.~\ref{Figure:LDOSFull}(a), the LDOS peak at the actual junction atom is also a bit smaller because the Majorana fermion in fact has its highest density on the neighboring sites.

\section{Intragap states}
\begin{figure}[tbh]
\includegraphics[width=225pt]{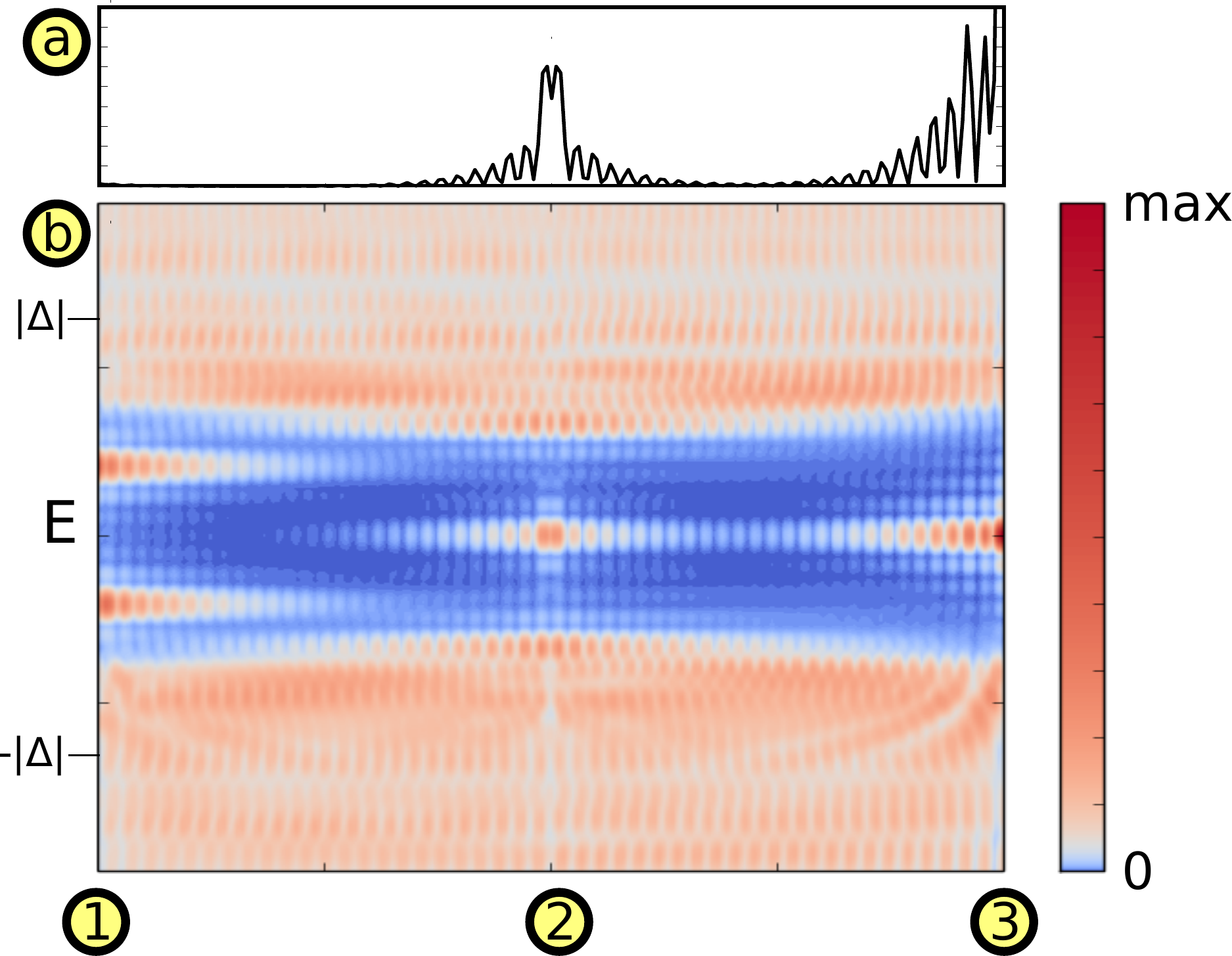}
\caption{LDOS along the line joining points 1 and 3, at 
$E = 0$ (a) and as function of energy (b).
White ripples along the energy direction in the otherwise empty region (blue) are due to the truncation of the Chebyshev expansion, and not due to additional states, and can be understood as Gibbs oscillations. They are only visible due to a logarithmic color scale (cp.~to Fig.~\ref{Figure:LDOS} plotted without logarithmic scale). Same parameters as for Fig.~\ref{Figure:LDOS}}.
\label{Figure:LDOSFull}
\end{figure}

Having seen that Majorana fermions appear only at odd junctions, we move on to discuss the other intragap states. For odd junctions they determine the excitation gap, which is important if the Majorana fermions are to be used for storing information. Further, the intra gap states at even junctions sets the energy scale needed for resolving the difference between even and odd junctions, important for providing evidence of Majorana fermions.
To study all types of intragap states we plot in Fig.~\ref{Figure:LDOSFull} the LDOS along the whole wire segment from point 1 to 3, spanning, wire-end, 2-, and 3-wire junctions. There is a continuum of intragap states along the whole wire segment, stretching about halfway into the gap $|\Delta|$ and with relatively little variation in the band bottom energy.
There exists also at the even-junction two intragap states slightly below this continuum, but their energies are not notably separated from the continuum band bottom along the wire segment as a whole. These states are also well-localized, revealing them being strongly hybridized wire-end points Majorana states. 
To thoroughly investigate the behavior of the intragap states over a wide parameter range, we also study the LDOS at both the even junction 1 and the odd junction 2 as a function of varying Rashba spin-orbit interaction and magnetic impurity strength. The variation with the other two parameters, chemical potential and superconducting order parameter, is much simpler and reported in the Supplementary material.

\section{Rashba spin-orbit interaction}
\begin{figure}[tbh]
\includegraphics[width=225pt]{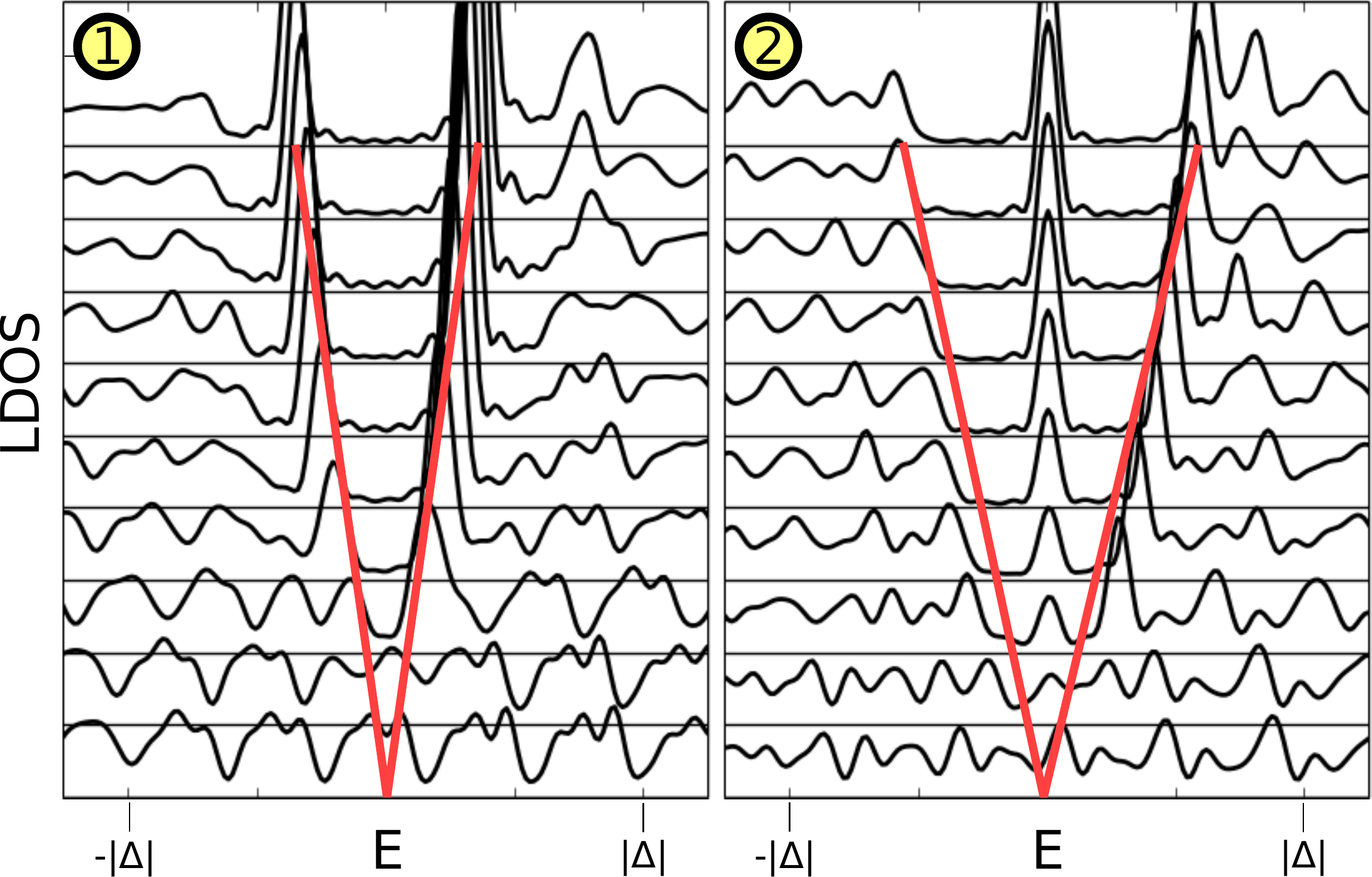}
\caption{LDOS at sites 1 and 2 for $\mu = -4, \Delta = 0.1$, $V_z = 1.5$ and $\alpha \in [0, 0.45]$ in steps of $0.05$. Lines are drawn with increasing offset with $\alpha = 0$ for the bottom line. Guidelines (red) indicate the lowest energy of the intragap states.}
\label{Figure:LDOS_alpha_junction_3}
\end{figure}

In Fig.~\ref{Figure:LDOS_alpha_junction_3} we plot the LDOS at sites 1 and 2 for a sequence of different spin-orbit coupling strengths $\alpha$. It is clear that the energy gap depends fairly linearly on the strength of the Rashba spin-orbit interaction. 
It is also clear that the Majorana fermions at the odd junction push the intragap states to notably higher energies, creating a larger excitation gap. For the even junction the gap is somewhat smaller. Notably, however, only for strong Rashba coupling there exists detectable states below the continuum.

To understand why the Rashba spin-orbit interaction opens up an energy gap it is instructive to consider what happens in the absence of Rashba spin-orbit interaction, and for small magnetic impurity strengths.
However, in the absence of both Rashba spin-orbit interaction and Zeeman spin-splitting, the chemical potential natural for studying topological superconductivity corresponds to a Fermi level close to band edge of the normal state Hamiltonian.
To remedy this problem and be able to study what happens as one term after another is turned on, we for a moment shift the chemical potential back firmly into the band of the normal state Hamiltonian, choosing $\mu = -2$.
The resulting model is not in the topologically non-trivial phase, but the intragap bands are a very general feature.

\begin{figure}
\includegraphics[width=220pt]{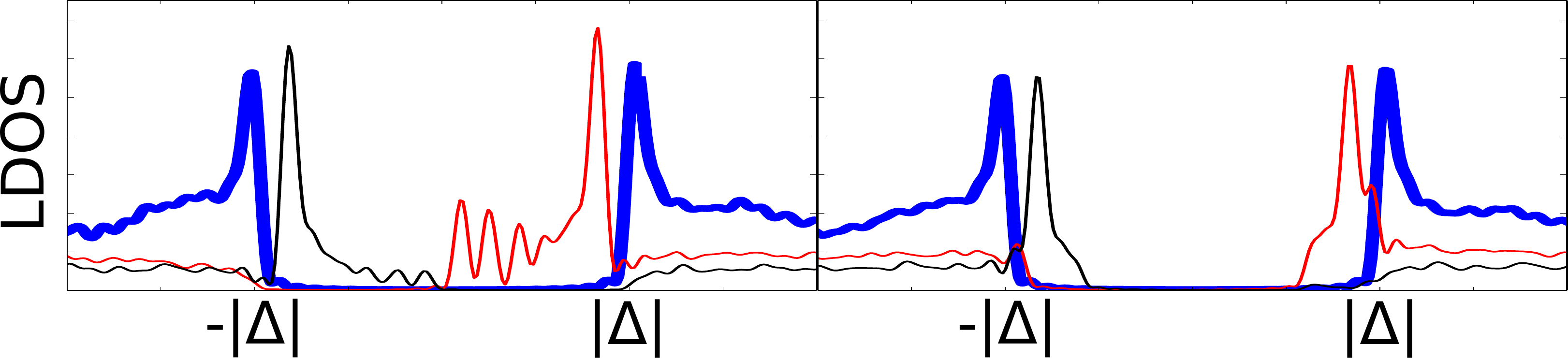}
\caption{Bulk LDOS for a conventional $s$-wave superconductor (thick blue line) with $\mu = -2$, as well as spin polarized LDOS for spin-up (red) and spin-down (black) electrons in the middle of a 200 sites long ferromagnetic wire with $V_z = 0.3$, embedded in the $s$-wave superconductor 400 x 201 sites large, for $\alpha = 0$ (left panel) and $\alpha = 0.3$ (right panel).}
\label{Figure:LDOSNormalWire}
\end{figure}

It is well known that a single magnetic impurity in an $s$-wave superconductor gives rise to localized Yu-Shiba-Rusinov (YSR) states inside the superconducting gap~\cite{RevModPhys.78.373, ActaPhysSin.21.75, ProgTheorPhys.40.435, JETPLett.9.85, PhysRevLett.115.116602}.
A chain of such impurities therefore give rise to what can be dubbed YSR bands.
In the left panel of Fig.~\ref{Figure:LDOSNormalWire} we plot the spin-polarized LDOS in the middle of a long ferromagnetic impurity wire embedded in an $s$-wave superconductor without Rashba coupling, as well as the bulk LDOS in the absence of impurities.
It is clear that the YSR bands are strongly spin-polarized, with spin-up electrons entering the energy gap from above, and spin down electrons entering from below, since the Zeeman spin-splitting term favors spin-up electrons in Eq.~\ref{Equation:Tight_binding_Hamiltonian}.
With increasing Zeeman spin-splitting, the intragap bands eventually starts to overlap, but the spin-up and down states are still independent of each other.
However, as soon as spin-orbit interaction is turned on, the two bands are coupled to each other and a gap opens up, as is seen in the right panel of Fig.~\ref{Figure:LDOSNormalWire}.
%

The opening of an energy gap facilitated by the spin-orbit term can also be understood from the band structure of a 1D ferromagnetic and superconducting wire. As discussed in the Supplementary material, the spin-orbit interaction only opens a gap in the presence of a superconducting order parameter.
In particular, this explains why the spin-orbit interaction pushes the YSR bands away from the Fermi level.
The Rashba term itself has no preference for hybridizing energy levels more strongly around the Fermi level, but superconductivity clearly has.

\section{Zeeman spin-splitting}
\begin{figure}
\includegraphics[width=220pt]{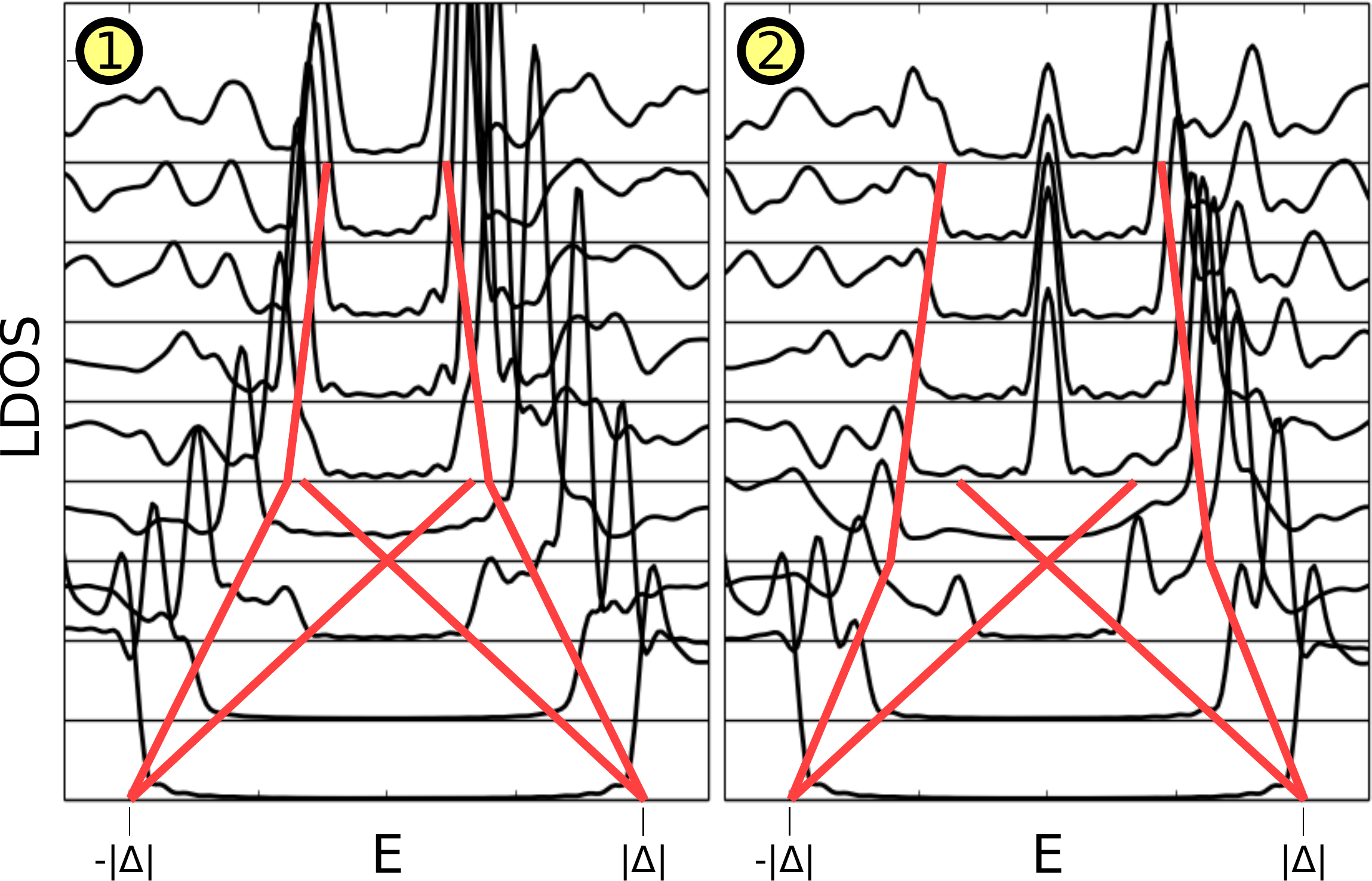}
\caption{LDOS at sites 1 and 2 for $\mu = -4, \alpha = 0.3, \Delta = 0.1$, and $V_z \in [0, 2]$ in steps of $0.25$. Lines are drawn with increasing offset with $V_z = 0$ for the bottom line. Guidelines (red) indicate the lowest energies of the intragap states.}
\label{Figure:LDOS_Vz_junction_3}
\end{figure}

We have already seen how the Zeeman spin-splitting in the absence of Rashba spin-orbit interaction affects the intragap states by pulling them through the gap.
Next we consider the effect of the Zeeman spin-splitting in the presence of all other terms.
In Fig.~\ref{Figure:LDOS_Vz_junction_3} we plot the LDOS at sites 1 and 2 for a sequence of different $V_z$.
The behavior of the LDOS is notably more complicated than for varying Rashba interaction.
This can be understood as a consequence of the dependence of the band structure on the Zeeman spin-splitting and the topological phase transition that the system goes through as $V_z$ is varied.
The topological phase transition occurs as the band gap closes at $k=0$ (along the wire), see Supplementary material. However, far away from the topological phase transition the gap is smallest at finite $k$, where it is instead opened by the Rashba spin-orbit interaction, see Supplementary material.
Both of these band edges therefore gives rise to signatures in the LDOS.

As indicated by the cross shaped guide lines in Fig.~\ref{Figure:LDOS_Vz_junction_3}, a rapid decrease of the energy gap first occurs as the Zeeman spin-splitting is turned on, until the topological phase transition occurs at the crossing point.
After the phase transition, this gap at $k=0$ increases again and eventually becomes too large to give any clear signature in the LDOS, as indicated by the terminated cross shaped lines.
In addition, another set of intragap states can also be identified, which follows the first type of intragap states but at a slower rate.
However, in contrast to the first type of intragap states, these states continues to move into the gap also after the topological phase transition, albeit at an even slower rate (see kink approximately at topological phase transition).
We also note that Fig.~\ref{Figure:LDOS_Vz_junction_3} eliminates any doubt regarding whether the zero energy state at the odd-junction at site 2 truly is a Majorana fermion, as it is seen to appear once the topological phase transition has occurred.
Figure~\ref{Figure:LDOS_Vz_junction_3} also displays the clear difference between intragap states between even- and odd-junctions for a very wide range of magnetic impurity strengths.

\section{Conclusions}
We have studied a wire network of ferromagnetic impurities on the surface of an $s$-wave superconductor with strong Rashba spin-orbit interaction.
We find that zero-energy Majorana fermions not only appear at the end points of individual wires, but also occur at junctions with an odd number of wire segments, while no low-energy states are present in junctions with an even number of wire segments.
This even-odd effect for the appearance of zero-energy states is a unique consequence of their Majorana character. Both construction and measurements of a wire network should be possible using existing scanning tunneling technology and would provide evidence for Majorana fermions much stronger than measurements on a single wire.
Furthermore, we find that the energy gap protecting the Majorana fermions in odd-junctions and fully gapping even-junctions, increases notably with larger Rashba spin-orbit coupling. By applying an electric field it might be possible to experimentally tune the strength of the surface Rashba spin-orbit coupling. Varying the magnetic impurity spin-splitting primarily tunes the topological phase, but has limited influence inside the topological phase.

\section{Acknowledgement}
This work was supported by the Swedish Research Council (Vetenskapsr\aa det), the Swedish Foundation for Strategic Research (SSF), the G\"oran Gustafsson Foundation, and the Wallenberg Academy Fellows program.

\clearpage

\noindent {\bf Supplementary material}
\\
\section{Chemical potential}
\label{Appendix:Chemical_potential}
\begin{figure}[h]
\includegraphics[width=235pt]{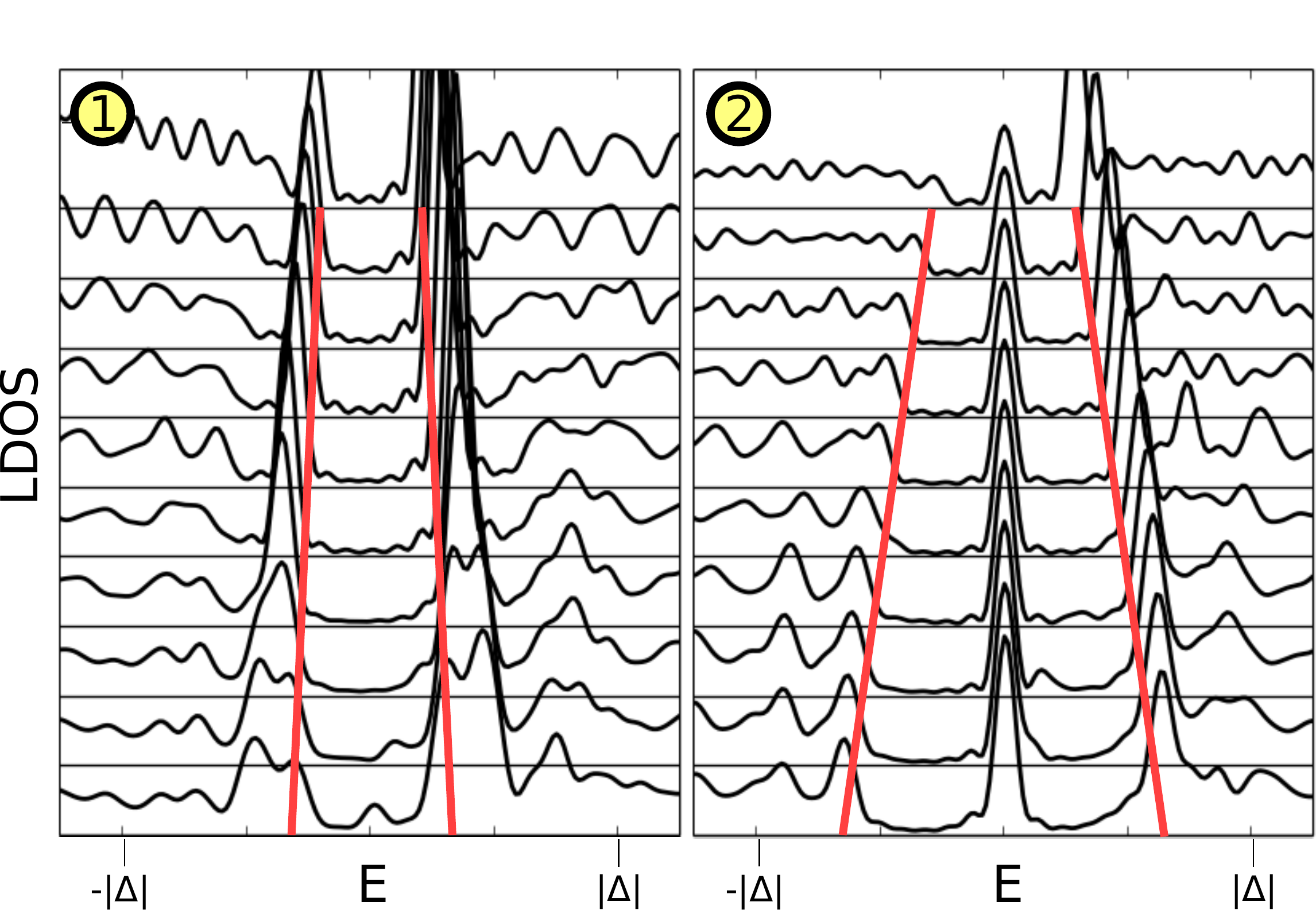}
\caption{LDOS at sites 1 and 2 for $\alpha = 0.3, \Delta = 0.1$, $V_z = 1.5$ and $\mu \in [-4.4, -3.5]$ in steps of $0.1$. Lines are drawn with increasing offset with $|\mu| = 3.5$ for the bottom line. Guidelines (red) indicate the lowest energy of the intragap states.}
\label{Figure:LDOS_mu_junction_3}
\end{figure}

In Fig.~\ref{Figure:LDOS_mu_junction_3} we plot the behavior of the LDOS at sites 1 and 2 as a function of the chemical potential $\mu$ moving towards the band edge. As seen, the energy gap can be seen to decrease smoothly as a function of $|\mu|$.
This can be understood as a consequence of the normal state DOS at the Fermi level being continuously decreased as the chemical potential approaches the band edge.

\section{Order parameter}
\label{Appendix:Order_parameter}
\begin{figure}[h]
\includegraphics[width=235pt]{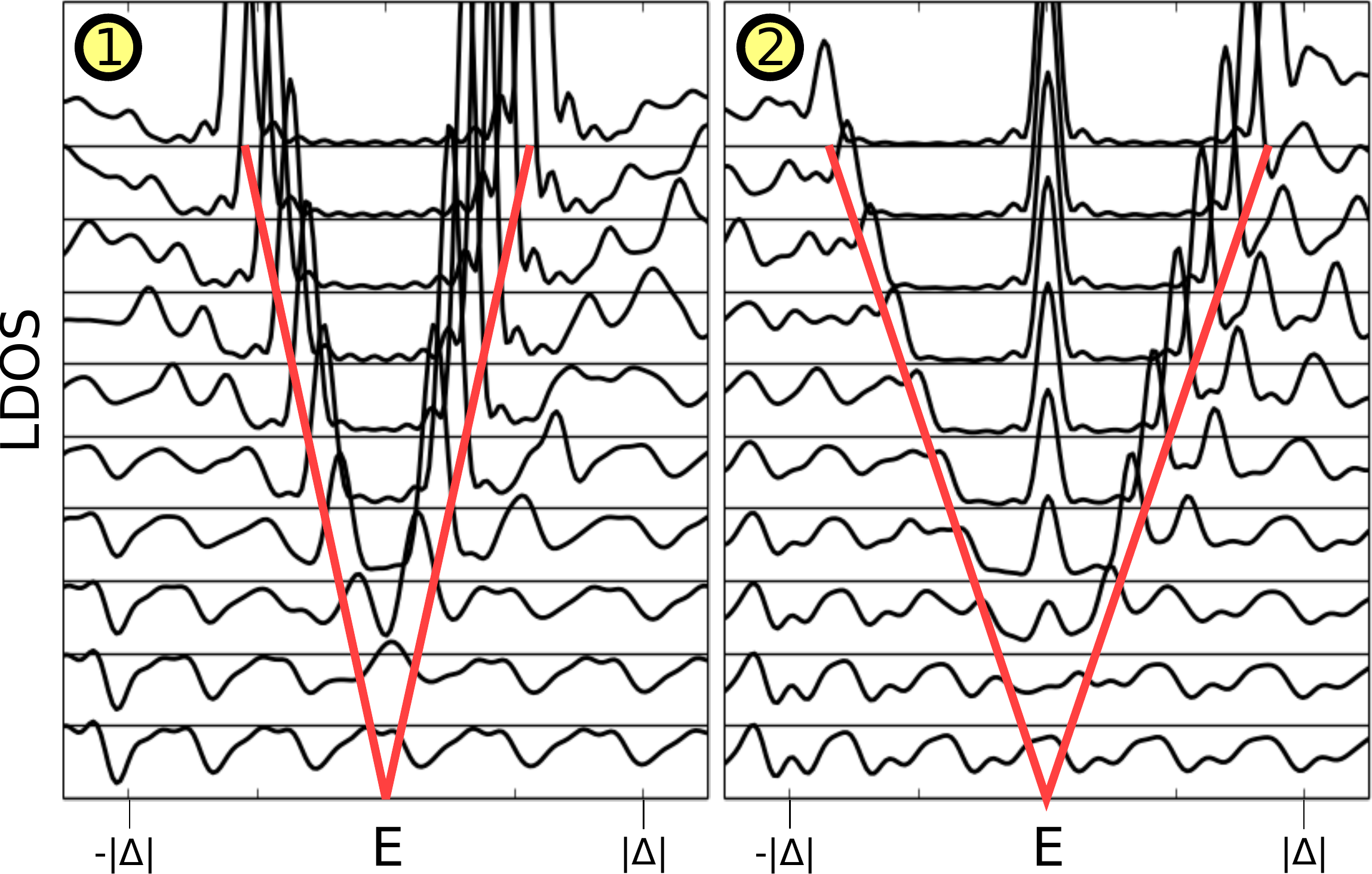}
\caption{LDOS at site 1 and 2 for $\mu = -4, \alpha = 0.3, \Delta = 0.1$, $V_z = 1.5$ and $\Delta \in [0, 0.18]$ in steps of $0.02$. Lines are drawn with increasing offset with $|\Delta| = 0$ for the bottom line. Guidelines (red) indicate the lowest energy of the intragap states.}
\label{Figure:LDOS_Delta_junction}
\end{figure}

In Fig.~\ref{Figure:LDOS_Delta_junction} we plot the behavior of the LDOS at site 1 and 2 as a function of the size of the superconducting order parameter $\Delta$.
The band gap can be seen to increase smoothly as a function of $\Delta$.

\section{Band-structure}
\label{Appendix:Band_structure}
\begin{figure}
\includegraphics[width=235pt]{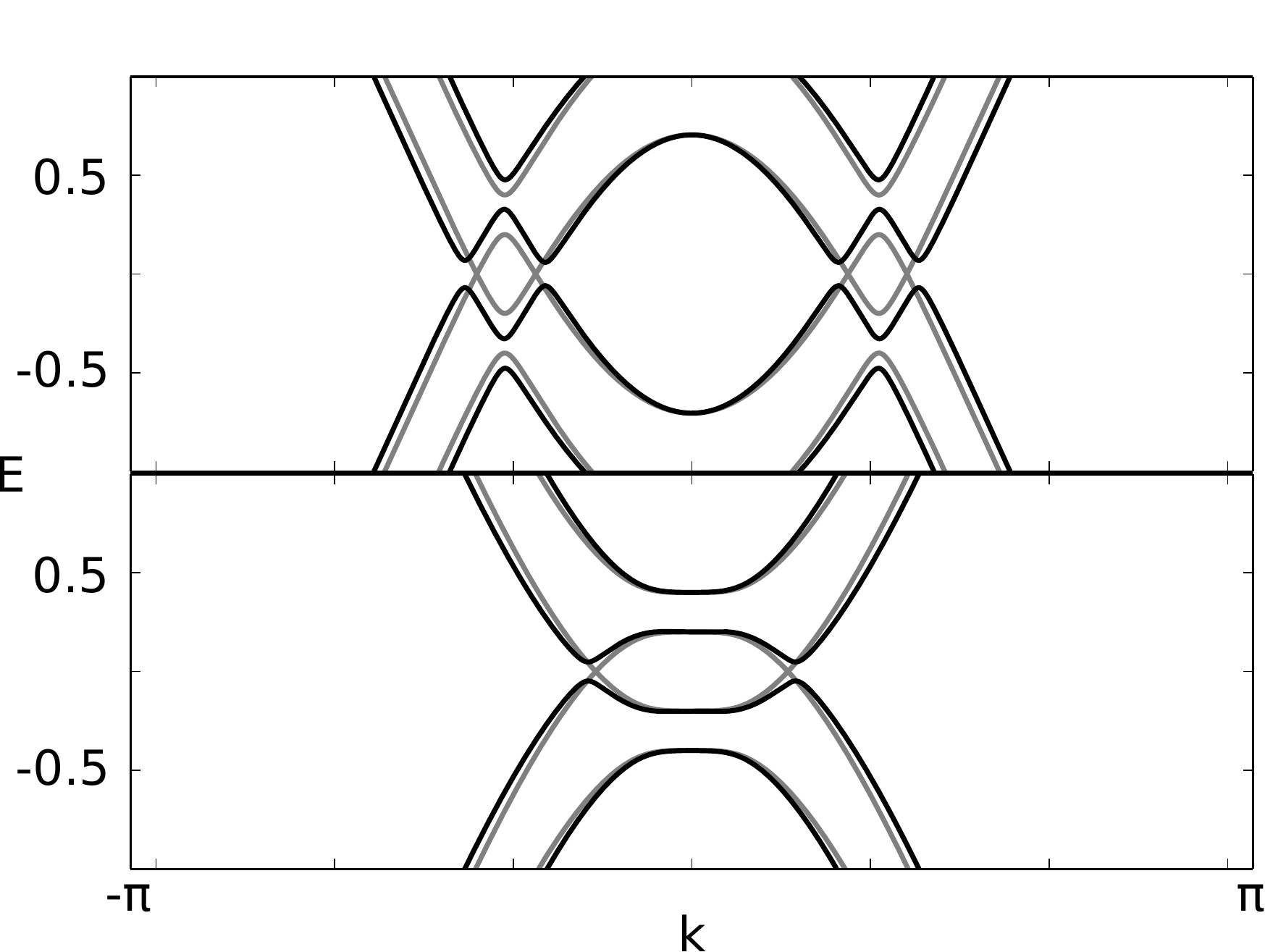}
\caption{Band structures of a 1D ferromagnetic and superconducting wire with $V_z = 0.3$, and $\Delta = 0.1$ for $\alpha = 0$ (grey lines) and $\alpha = 0.3$ (black lines). (Top) trivial phase with $\mu = -1$. (Bottom) non-trivial phase with $\mu = -2$.}
\label{Figure:BandStructure}
\end{figure}

Here we will demonstrate how the Rashba spin-orbit interaction opens up an excitation gap in a superconducting and ferromagnetic wire by considering a 1D bulk Hamiltonian for such a system in the continuum limit:
\begin{align}
	\mathcal{H}_{1D} =& \left[\begin{array}{cccc}
			\epsilon(k) - V_z & -\mathcal{L}(k) & 0 & \Delta\\
			\mathcal{L}(k) & \epsilon(k) + V_z & -\Delta & 0\\
			0 & -\Delta^{*} & -\epsilon(k) + V_z & \mathcal{L}(k)\\
			\Delta^{*} & 0 & -\mathcal{L}(k) & -\epsilon(k) - V_z
		\end{array}\right],
\end{align}
where $\epsilon(k) = -2t\cos(k) - \mu$ and $\mathcal{L}(k) = i\alpha\sin(k)$.
The energy spectrum for this Hamiltonian is~\cite{PhysRevB.82.134521, PhysRevB.91.214514}:
\begin{widetext}
\begin{align}
	E(k) = \pm\sqrt{\epsilon^2(k) + \mathcal{L}^2(k) + V_z^2 + |\Delta|^2 \pm 2\sqrt{\epsilon^2(k)\mathcal{L}^2(k) + \left[\epsilon^2(k) + |\Delta|^2\right]V_z^2}}.
\end{align}
\end{widetext}
In Fig.~\ref{Figure:BandStructure} we plot this band structure for both $\alpha = 0$ and $\alpha = 0.3$.
In the top panel the chemical potential is set to $\mu = -1$, corresponding to the trivial phase, while in the bottom panel $\mu = -2$, corresponding to the topologically non-trivial phase. The topological phase transition takes place when the bands invert at $k = 0$.
(Note that the band edge of the 1D model is around $\mu = -2$ rather than around $\mu = -4$ as in the 2D model.)
%

As clearly seen, the introduction of the Rashba spin-orbit interaction opens an energy gap in the band structure at a finite $k$-value, for both the trivial and non-trivial topological phases.
In fact, this gap opening should be understood as an interplay between the Rashba spin-orbit interaction and superconductivity, since the introduction of a Rashba interaction fails to open a gap in the absence of superconductivity.
To understand why Rashba spin-orbit interaction and superconductivity are both needed, we note that in the absence of superconductivity, the Rashba-spin orbit interaction couples the two Zeeman split electron band parabolas to each other, and likewise for the holes.
However, these do not cross each other and a gap opening can therefore not take place.
Similarly, in the absence of Rashba spin-orbit interaction, the superconducting order parameter couples the spin-up electron band to the spin-down hole band, and vice versa.
For a finite Zeeman spin-splitting these pair of bands however only cross each other away from the Fermi level (if at all), and explains the four gap openings that appear away from the Fermi level for the grey bands in the top panel of Fig.~\ref{Figure:BandStructure}.
However, when both terms are present at the same time, all four band couple to each other.
The Rashba spin-orbit interaction mixes the spin-character of the electron and hole bands, allowing the superconducting order parameter to gap any band crossing.
In particular, this explains why the addition of a spin-orbit interaction leads to an energy gap opening around the Fermi level, even though there is nothing special about the Fermi level for the Rashba spin-orbit interaction in itself.

\end{document}